\documentclass[aps,prb,twocolumn]{revtex4-2}
\usepackage{setspace, parskip, xcolor, breakcites, hyphenat, graphicx, epstopdf, amsmath, amssymb, amsfonts, array,
	verbatim, url, subcaption, booktabs, multirow, etoolbox,siunitx, etoolbox, tikz,stackengine, textcomp,
	gensymb, siunitx,tabularx, array}
\def\stripzero#1{\expandafter\stripzerohelp#1}

\def\stripzerohelp#1{\ifx 0#1\expandafter\stripzerohelp\else#1\fi}
\newcolumntype{H}{>{\setbox0=\hbox\bgroup}c<{\egroup}@{}}
\usepackage[toc,page,titletoc]{appendix}
\usepackage{dcolumn}        % alignment in tables
\usepackage{bm}             % bold math
\usepackage[version=4]{mhchem}
\usepackage{siunitx}        % consistent units: \SI{5.41}{\angstrom}
\definecolor{srm}{HTML}{034da1}
\definecolor{cof}{RGB}{219,144,71}
\definecolor{pur}{RGB}{186,146,162}
\definecolor{greeo}{RGB}{91,173,69}
\definecolor{greet}{RGB}{52,111,72}
\definecolor{mplc0}{HTML}{1f77b4}
\definecolor{mplc1}{HTML}{ff7f0e}
\definecolor{mplc2}{HTML}{2e7d32}
\definecolor{mplc3}{HTML}{d32f2f}
\definecolor{mplc4}{HTML}{9467bd}
\definecolor{mplc5}{HTML}{8c564b}
\definecolor{mplc6}{HTML}{e377c2}
\definecolor{mplc7}{HTML}{7f7f7f}
\definecolor{mplc8}{HTML}{bcbd22}
\definecolor{mplc9}{HTML}{17becf}
\definecolor{Astral}{HTML}{1F77B4}
\definecolor{BG80}{HTML}{37474f}
\definecolor{Green}{HTML}{4CA540}
\definecolor{Red}{HTML}{F44336}
\usepackage[colorlinks=true, linkcolor=mplc0, urlcolor=mplc3, filecolor=mplc0, citecolor=mplc2,
  pdfstartview=FitV, pdftitle={}, pdfauthor={}, pdfsubject={}, pdfkeywords={}, pdfpagemode={},
  bookmarksopen=true, breaklinks]{hyperref}

\graphicspath{{./Figures/}}
\newcommand{\CeO}{\ce{CeO2}}
\newcommand{\CeIII}{Ce$^{3+}$}

\newcommand{\eV}{\,\text{eV}}

\makeatletter
\def\@email#1#2{%
	\endgroup
	\patchcmd{\titleblock@produce}
	{\frontmatter@RRAPformat}
	{\frontmatter@RRAPformat{\produce@RRAP{*#1\href{mailto:#2}{#2}}}\frontmatter@RRAPformat}
	{}{}
}%
\makeatother
\begin{document}

\title{Hyperfine Structure and Exchange Coupling of Vacancy-Induced \ce{Ce^3+} Spin Centers in Nuclear-Spin-Dilute \ce{CeO2}}
\author{T.M. Chithresh}
\author{Rudra Banerjee}\email{rudrab@srmist.edu.in}
\affiliation{Department of Physics and Nanotechnology, SRM Institute of Science and Technology, Kattankulathur, Tamil Nadu, 603203, India}

\begin{abstract}
	Oxygen vacancies in cerium dioxide (\ce{CeO2}) donate electrons that localize as \ce{Ce^3+} ($4f^1$, $S=1/2$) small polarons, creating rare-earth spin centers through native defect chemistry rather than implantation or extrinsic doping. We investigate the magnetic environment of these centers using first-principles PBE$+U$ calculations with a linear-response Hubbard parameter ($U=\SI{5.8382}{\eV}$), hyperfine tensors from the all-electron reconstruction of the projector-augmented-wave method, and Korringa--Kohn--Rostoker exchange calculations within the coherent-potential approximation. Four vacancy configurations spanning concentrations from $3.125\%$ to $12.5\%$ are considered. A distinctive feature of the host follows from cerium isotopics: all naturally occurring cerium isotopes possess nuclear spin $I=0$, eliminating on-site hyperfine interactions at the \ce{Ce^3+} center and leaving the nuclear-spin bath entirely on the oxygen sublattice, whose sole magnetic isotope, \ce{^17O} ($I=5/2$), occurs at $0.038\%$ natural abundance. The resulting \ce{^17O} hyperfine landscape consists of a small number of strongly coupled, nearly axial first-shell nuclei with contact couplings reaching ${\sim}6$~MHz, surrounded by a weakly coupled and strongly anisotropic outer shell. These tensors define experimentally accessible signatures for \ce{^17O} ESEEM and HYSCORE measurements and provide the microscopic hyperfine parameters required for cluster-correlation-expansion calculations of spin coherence. Exchange interactions between neighboring polarons are weak and oxygen-mediated, leaving the vacancy-generated spins largely independent over the concentration range considered. Together, these results establish oxygen-deficient \ce{CeO2} as a chemically generated and intrinsically nuclear-spin-dilute host for rare-earth spin centers, and provide the first-principles magnetic parameters needed to assess their coherence properties.
\end{abstract}
% 71.55.-i  Impurity and defect levels
% 75.10.Dg  Crystal-field theory and spin Hamiltonians
% 03.67.Lx  Quantum computation architectures and implementations
% 71.15.Mb  Density functional theory (DFT)

% \keywords{cerium oxide, oxygen vacancy, Ce$^{3+}$, hyperfine interaction,
% spin qubit, DFT$+U$, polaron, quantum information}

\maketitle
\section{Introduction}\label{sec:intro}

Solid-state spin defects have become a leading platform for quantum technologies, functioning as qubits, single-photon sources, and nanoscale sensors~\cite{Wolfowicz2021, DeLeon2021}. Their appeal lies in combining atom-like quantum coherence with the scalability of the host crystal, and in favorable materials this coherence persists to room temperature. The defect's performance is governed in large part by its spin coherence time $T_2$~\cite{Wolfowicz2021}, the interval over which the spin retains quantum phase information before environmental fluctuations render the state classical. In wide-gap insulators free of paramagnetic impurities, $T_2$ is set by the fluctuating magnetic field of the surrounding nuclear-spin bath, which couples to the electron spin through the hyperfine interaction~\cite{deSousa2003}. The materials-design problem is therefore to identify hosts that simultaneously provide a wide band gap, a high Debye temperature, a sparse nuclear-spin environment, and---critically for scalable architectures---a reproducible, deterministic means of placing the spin centers~\cite{DeLeon2021}.

The nitrogen-vacancy (NV) center in diamond is the archetype against which candidate systems are measured: its spin-triplet ground state, large zero-field splitting, and near-spin-free $^{12}$C host ($I=0$ at $98.9\%$ natural abundance) yield millisecond coherence at ambient conditions~\cite{Wolfowicz2021, Doherty2013}, extending to $\sim$2\,ms under isotopic purification~\cite{Balasubramanian2009}. Two practical constraints accompany this performance: isotopic purification is costly at scale, and deterministic placement still relies on ion implantation with limited spatial yield. These have driven a broad search for alternatives---among silicon-carbide divacancies~\cite{Koehl2011, Christle2015}, rare-earth-doped insulators~\cite{Zhong2015}, and wide-gap oxides and nitrides~\cite{Ivady2018, Lee2022, Kanai2022}---whose most successful members share a defining feature: the spin center is generated and positioned by the intrinsic chemistry of the host rather than implanted. The present work is motivated by this design principle.

Rare-earth ions in solids supply a second, complementary benchmark. The record coherence among solid-state spins---six hours, for the $^{151}$Eu$^{3+}$ nuclear spin in Y$_2$SiO$_5$ under optimized optical and magnetic-field conditions~\cite{Zhong2015}---belongs to this family, and its Kramers members reach second-scale hyperfine coherence, as in isotopically purified $^{167}$Er$^{3+}$:Y$_2$SiO$_5$ ($4f^{11}$)~\cite{Rancic2018}. The mechanism is the screening of the $4f$ shell by the filled $5s^2 5p^6$ outer shells, which suppresses crystal-field broadening and spin--phonon coupling relative to $d$-or $p$-electron defect spins~\cite{Thiel2011}. A rare-earth center inherits this shielded-$4f$ advantage in principle; what differs between hosts is whether the chemistry places the ion at reproducible sites without doping-induced disorder, and what local nuclear-spin environment surrounds it.

Cerium dioxide (\CeO{}, ceria) is an unusually attractive yet underexplored realization of this idea. It is a fluorite-structure insulator with a band gap of $\sim$3.3\,eV~\cite{Han2016, Wuilloud1984} and a high Debye temperature, $\sim$450\,K~\cite{Morrison2019}. Removing a lattice oxygen donates two electrons that localize as Ce$^{3+}$ ($4f^1$, $S=1/2$) small polarons on cerium sites near the vacancy~\cite{Skorodumova2002, Sun2017, Murgida2014, Ganduglia2009}, so the spin center is created by native defect chemistry---through controlled reduction---rather than by implantation. The resulting Ce$^{3+}$ is a rare-earth Kramers ion that inherits the shielded-$4f$ character of Er$^{3+}$:Y$_2$SiO$_5$, but in a host where the center arises from intrinsic stoichiometry rather than substitutional doping. Equally important, the fluorite lattice admits several crystallographically distinct vacancy arrangements at a given stoichiometry~\cite{Murgida2014, Han2016}, making vacancy geometry an explicit control parameter: it sets the spacing and relative orientation of the polarons and thereby tunes both their mutual exchange and the hyperfine environment that governs $T_2$. A spin center that is chemically generated and structurally tunable in this way occupies a different design space from the implanted point defects that dominate the qubit literature.

Ceria carries a further, decisive advantage that follows directly from cerium isotopics and that reframes what must be computed.
Every stable cerium isotope has nuclear spin $I=0$, so the cerium sublattice is intrinsically nuclear-spin-free at natural
abundance---the same condition that the spin-free $^{12}$C host of NV-diamond attains only through isotopic purification. The
Ce$^{3+}$ electron spin therefore has no on-site hyperfine partner, and the bath that limits $T_2$ is carried entirely by oxygen,
whose sole magnetic isotope is $^{17}$O ($I=5/2$)\cite{Stone2005} at a natural abundance of just $0.038\%$~\cite{Meija2016}. The
center thus sits in an exceptionally dilute bath whose strength is, moreover, tunable through $^{17}$O enrichment---a structural
analogue of the $^{12}$C--$^{13}$C control exploited in diamond\cite{Balasubramanian2009}. The same host nuclear-spin environment has separately been exploited for extrinsically doped Er$^{3+}$:CeO$_2$ telecom spin qubits~\cite{Zhang2024,Seth_2025}; the vacancy-induced center considered here differs in carrying no on-site nuclear moment at the spin-bearing ion itself; unlike \ce{^167Er} ($I=7/2$), every Ce isotope has $I=0$. What downstream coherence modeling requires is therefore not an on-site cation coupling but the distribution of $^{17}$O hyperfine tensors around each polaron.
What downstream coherence modeling requires is therefore not an on-site cation coupling but the distribution of $^{17}$O hyperfine tensors around each polaron.

This work computes the static spin-Hamiltonian inputs for that modeling---the $^{17}$O hyperfine tensors and the inter-polaron
exchange---within collinear, scalar-relativistic DFT$+U$, and two features of this level of theory should be stated at the
outset. The first is spin--orbit coupling. The physical ground state of $4f^1$ Ce$^{3+}$ is the $^2F_{5/2}$ multiplet, whose
lowest crystal-field Kramers doublet defines the effective spin-$1/2$ qubit and carries an anisotropic $g$
tensor\cite{AbragamBleaney1970}; the collinear calculation omits this splitting by construction. What it retains is the quantity
the bath depends on: the Fermi-contact and dipolar couplings at the surrounding oxygens are fixed by the transferred $s$ and $p$
spin density, which is governed by the scalar spin distribution and is captured here. The $g$-tensor anisotropy and the
projection of $^2F_{5/2}$ onto the effective doublet---the corrections that map these tensors onto the measured spin-$1/2$
Hamiltonian---are the subject of a companion all-electron study, for which the present tensors are the input. The second feature
is concentration: the supercells that expose polaron--polaron coupling sit at reduced-ceria vacancy concentrations
($3.125$--$12.5\%$), far above any realistic single-qubit density. Because the hyperfine tensor of a given $^{17}$O nucleus is a
near-field property of the nearest polaron's spin density, the per-polaron bath is a local quantity transferable to the
isolated-defect limit, whereas the multi-vacancy cells serve specifically to probe how proximity perturbs the exchange. We make
no claim to a functioning qubit: optical initialization, single-spin readout, and microwave control remain open problems for
ceria. The hyperfine predictions are, however, directly testable, since electron-spin-echo envelope modulation
(ESEEM)\cite{Mims1972} and its two-dimensional extension HYSCORE\cite{Hoefer1986} resolve exactly the weak-to-moderate $^{17}$O
ligand couplings reported here\cite{Schweiger2001}.

Three results follow. First, the localized $4f^1$ spin density coexists with a silent cation: the polaron is fully formed yet
contributes no on-site hyperfine channel, so the dilute $^{17}$O sublattice is the entire bath---realizing without isotopic
purification the spin-free-host condition that underlies NV-diamond coherence.
Second, the $^{17}$O tensors separate into a small set of strongly coupled, nearly axial sites in the first coordination shell of each polaron (contact couplings up to $\sim$6\,MHz) and a surrounding shell of weakly coupled but strongly anisotropic sites; together they form the complete first-principles bath input for a cluster-correlation-expansion treatment\cite{Yang2008, Witzel2006}, and their anisotropy carries a geometry-dependent pattern that defines a target for $^{17}$O ESEEM/HYSCORE. Third, the inter-polaron exchange is weak (sub-meV) and is mediated through the oxygen sublattice rather than by direct Ce--Ce overlap; its sign is not robustly fixed at the present level---a mean-field (CPA) estimate favors a ferromagnetic tendency, while the supercell solutions are near-degenerate and magnetically compensated rather than sharply ordered---so the Ce$^{3+}$ moments are best described as magnetically dilute and near-independent. Together these establish \CeO{} as a chemically generated, structurally tunable, and intrinsically nuclear-spin-dilute host for Ce$^{3+}$ spin centers, and they supply the $^{17}$O dataset required for the coherence-time analysis to follow.

\section{Computational Methods}\label{sec:methods}
\subsection{Electronic Structure and Defect Models}\label{subsec:dft}

Spin-polarized, scalar-relativistic density-functional-theory calculations (collinear spin, with spin--orbit coupling omitted by design; Sec.~\ref{subsec:charge}) were performed with VASP~\cite{Kresse1996a, Kresse1996b} using the projector-augmented-wave (PAW) method~\cite{Blochl1994, Kresse1999} and the PBE exchange--correlation functional~\cite{Perdew1996}. The PBE PAW dataset for Ce carries the $5s$, $5p$, $6s$, $5d$, and $4f$ shells in valence (12 valence electrons), and that for O the $2s$ and $2p$ shells; strong on-site correlation in the Ce\,$4f$ manifold was treated with the rotationally invariant, simplified (Dudarev) $+U$ scheme~\cite{Anisimov1991, Dudarev1998} applied only to Ce\,$4f$. A plane-wave cutoff of \SI{700}{\eV}, and an electronic-convergence threshold of \SI{e-5}{\eV} were used, with Gaussian smearing of width \SI{0.05}{\eV}. These settings were applied identically to the pristine-cell, relaxation, and property (density-of-states and hyperfine) calculations, so that all reported comparisons are internally consistent.

\begin{figure}[h]
	\centering
	\includegraphics[width=.7\columnwidth]{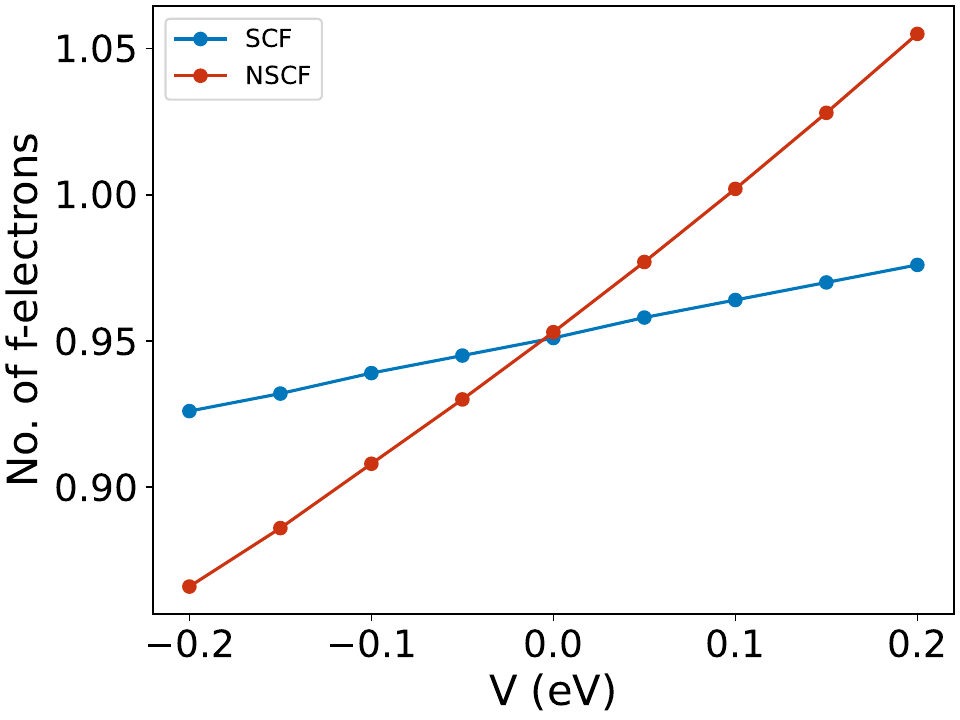}
	\caption{\label{fig:lr_u}Linear-response determination of the Hubbard $U$ for Ce\,$4f$ states in \CeO{}. The difference between the self-consistent (SCF) and non-self-consistent (NSCF) $4f$-occupancy responses to the applied perturbation $V$ yields $U = \SI{5.8382}{\eV}$.}
\end{figure}

The Hubbard parameter $U = \SI{5.8382}{\eV}$ was obtained from first principles by the linear-response method of Cococcioni and de Gironcoli~\cite{Cococcioni2005}, without empirical fitting, as the difference between the inverse non-interacting and interacting Ce\,$4f$ occupation responses to a localized potential perturbation applied on a single Ce site (perturbation strengths $-0.20$ to $+0.20$\,eV in \SI{0.05}{\eV} steps; Fig.~\ref{fig:lr_u}). The single value was held fixed across all configurations; it lies within the $5$--$6$\,eV range established for ceria~\cite{Murgida2014, Sun2017, Loschen2007} and localizes the \ce{Ce^3+} $4f^1$ polaron correctly (Sec.~\ref{subsec:bulk}). We use PBE$+U$ rather than a hybrid functional because reliable Fermi-contact couplings require dense reciprocal-space property evaluation on cells hosting up to eight polarons, for which hybrid exchange is prohibitive at these sizes; the central observable is the all-electron spin density at the nucleus, set by on-site $4f$ localization rather than by the fundamental gap, and is therefore largely insensitive to the residual gap underestimate of PBE$+U$~\cite{Ivady2018}.

Defect supercells were built from the relaxed $2\times2\times1$ conventional fluorite cell (48 atoms) by removing one to four oxygen atoms, spanning concentrations of $3.125\%$ to $12.5\%$ and, at $6.25\%$, contrasting clustered with dispersed vacancy pairs (Table~\ref{tab:systems}). We treat the charge-neutral oxygen vacancy: each removed O donates two electrons that localize as two \ce{Ce^3+} ($4f^1$) polarons, so every cell is electrically neutral and no charged-defect finite-size correction is required~\cite{Freysoldt2014}. All cells share the same base cell, fixing the defect--image separation across the series. Targeted polaron arrangements were stabilized by initializing the Ce moments on the chosen sites with the desired relative orientation, with no constraint on the total moment; the resulting localization was confirmed a posteriori from the site-projected $4f$ moments and the in-gap density of states (Sec.~\ref{subsec:charge}), guarding against the metastable delocalized solutions to which $+U$ ceria is prone~\cite{Murgida2014, Loschen2007}. Geometries were relaxed at the zone center, with full optimization of cell shape, volume, and internal coordinates, until residual forces fell below \SI{0.01}{\electronvolt\per\angstrom}; the relaxed lattice parameters are listed in Table~\ref{tab:systems}. Because the vacancy-induced $4f^1$ states are dispersionless gap states, all electronic, magnetic, and hyperfine properties were then evaluated in a separate self-consistent step on a dense $4\times4\times8$ $\Gamma$-centered mesh, which fully determines the occupied-state and contact-hyperfine quantities; only the relative orientation of the polaron moments---governed by sub-meV exchange (Sec.~\ref{subsec:exchange})---remains sensitive to sampling, and we accordingly report the magnetic solutions as converged self-consistent states rather than verified ground states.

\begin{table}[h]
	\caption{\label{tab:systems}Vacancy supercell configurations, all built on the same $2\times2\times1$ base cell. Lattice parameters follow full relaxation; $d_{V_c\text{--}\mathrm{Ce}^{3+}}$ is the mean vacancy--polaron distance.}
	\resizebox{\linewidth}{!}{
		\begin{tabular}{llccc}\toprule\toprule
			Label       & Description              & Conc.\ (\%) & $a$ (\AA) & $d_{V_c-\ce{Ce^{3+}}}$(\AA) \\\midrule
			Single      & 1 O removed, $[110]$     & 3.125       & 10.830    & 2.44                        \\
			Dbl-far     & 2 O, maximally separated & 6.25        & 10.870    & 2.42                        \\
			Dbl-close   & 2 O, adjacent            & 6.25        & 10.858    & 2.54                        \\
			Quad-closer & 4 O, clustered           & 12.5        & 10.971    & 2.56                        \\
			\bottomrule\bottomrule
		\end{tabular}
	}
\end{table}

The oxygen-vacancy formation energy per vacancy is
\begin{equation}
	E_\mathrm{f} = \frac{1}{n_V}\bigl[\, E(\text{CeO}_{2-x}) - E(\text{CeO}_2) + n_V\, \mu_\mathrm{O} \,\bigr],
	\label{eq:Ef}
\end{equation}
where $n_V$ is the number of vacancies, $E(\text{CeO}_{2-x})$ and $E(\text{CeO}_2)$ are the total energies of the defective and pristine supercells, and $\mu_\mathrm{O} = \tfrac{1}{2}E(\text{O}_2)$ is the oxygen chemical potential in the oxygen-rich (standard-state) limit. The GGA$+U$ description of the \ce{O2} molecule introduces a systematic offset in the absolute reference $\mu_\mathrm{O}$~\cite{Wang2006, Stevanovic2012}; because this offset is common to all configurations, it cancels in the relative formation energies and the clustering comparison on which our conclusions rest, and we accordingly interpret only relative formation energies.

\subsection{Hyperfine Tensors}\label{subsec:hfmethod}

The hyperfine tensor separates into an isotropic Fermi-contact term $A_\mathrm{iso}\propto\rho_s(\mathbf{R}_I)$, proportional to
the all-electron spin density at the nucleus, and a traceless dipolar term. The diagonalized principal values of the full tensor,
ordered $|A_{zz}|\geq|A_{xx}|\geq|A_{yy}|$, define the asymmetry $\eta=(A_{yy}-A_{xx})/A_{zz}$, which vanishes for an axially
symmetric coupling. Tensors were evaluated from the converged spin density of each relaxed cell within the PAW
formalism~\cite{Blochl2000, Yazyev2005, Szasz2013}. The contact term combines the smooth-grid (pseudo) spin density at the
nucleus with the on-site all-electron PAW reconstruction---which replaces the pseudo on-site partial-wave contribution and
restores the nuclear-cusp and nodal structure that the smooth density omits---and, in addition, the spin-polarization
contribution of the core states; the latter is non-negligible for the light \ce{^17O} nucleus (of order $20\%$ of the valence
contact term, and of opposite sign) and is included for quantitative couplings~\cite{Szasz2013}.
% CONFIRM(rudra): verify the reported A_iso INCLUDES the core term --- VASP's printed A_tot = A_pw - A_1PS + A_1AE EXCLUDES A_1c by default

Because every naturally abundant cerium isotope has nuclear spin $I=0$~\cite{Stone2005}, the cerium gyromagnetic ratio is identically zero and Ce carries no hyperfine coupling; the sole magnetic nucleus of the host is \ce{^17O} ($I=5/2$), and all reported couplings use its gyromagnetic ratio $\gamma_n/2\pi=\SI{-5.7735}{\mega\hertz\per\tesla}$~\cite{Stone2005}. The \ce{^17O} couplings are therefore the only hyperfine observable.

Hyperfine tensors computed under periodic boundary conditions are contaminated by periodic images for the weak, long-range couplings, where the spin density is small and image overlap dominates~\cite{Takacs2024}. This does not affect our conclusions: the quantitatively used couplings are the strong, first-shell \ce{^17O} interactions set by spin density within $\sim$1\,\AA{} of the nucleus---the most strongly converged regime---and the common base cell fixes the image separation across the series, so residual errors largely cancel in the concentration- and geometry-dependent trends. The weak, distant couplings enter only through their asymmetry distributions; their absolute strengths would require the real-space scheme of Ref.~\cite{Takacs2024} before use in spin-bath modeling.

\subsection{Inter-Polaron Exchange Coupling}\label{subsec:sprkkr}

Exchange constants were computed independently with the all-electron, fully relativistic Korringa--Kohn--Rostoker (KKR) multiple-scattering Green's-function method as implemented in SPR-KKR~\cite{Ebert2011}, at the PBE$+U$ level with the same $U=\SI{5.8382}{\eV}$ on Ce\,$4f$ (atomic-limit double counting) for consistency with the supercell calculations; the four-component relativistic treatment includes spin--orbit coupling at the heavy Ce site, complementing the scalar-relativistic supercell calculations. Vacancy disorder was treated within the coherent-potential approximation (CPA), each anion site being modeled as a statistical O/vacancy (empty-sphere) mixture at the corresponding concentration. Coherent-potential treatment of substitutional or vacancy disorder, in place of an explicit supercell representation, is established practice within the SPR-KKR framework~\cite{Kohler2016}. The CPA therefore returns the configurationally averaged pairwise exchange of a dilute, randomly distributed vacancy ensemble and does not resolve the close--far distinction of the ordered supercells---a scope limitation noted where the values are used (Sec.~\ref{subsec:exchange}). The averaging extends to the cerium valence itself: single-site CPA represents the cation sublattice by a single effective medium, and attempts to stabilize a Ce$^{3+}$/Ce$^{4+}$ contrast on that sublattice through charge- and spin-initialization relax uniformly to a single valence, the approximation having no mechanism to sustain a locally symmetry-broken, integer-occupation $4f^1$ state. The reported $J_{ij}$ are accordingly a valence- and configuration-averaged, dilute-limit estimate of the exchange scale; the site-specific Ce$^{3+}$ localization that defines the polarons is resolved by the supercell calculations (Sec.~\ref{subsec:charge}), not by the CPA. The Green's function was integrated on a semicircular energy contour of 30 energy points, using an angular-momentum cutoff $\ell_\mathrm{max}=3$ and a $1000$-point Brillouin-zone mesh. Pairwise parameters $J_{ij}$ of the classical Heisenberg Hamiltonian $H=-\sum_{i\neq j}J_{ij}\,\hat{\mathbf{e}}_i\!\cdot\!\hat{\mathbf{e}}_j$, with unit vectors $\hat{\mathbf{e}}_i$ along the local moments and $J_{ij}>0$ ferromagnetic, were obtained from the magnetic force theorem via infinitesimal rotations of the converged collinear moments~\cite{Liechtenstein1987}, summed within a real-space cluster of radius $2.4\,a_0$, where $a_0$ denotes the cubic lattice constant (also the unit of the pair separations in Table~\ref{tab:jij} and Fig.~\ref{fig:jij}).

\section{Results and Discussion}\label{sec:results}

\subsection{Pristine \CeO: Reference Electronic Structure}\label{subsec:bulk}

We first fix the electronic structure of pristine \ce{CeO2} against which the defect states are referenced. \ce{CeO2} is a fluorite charge-transfer insulator with an O\,$2p$ valence band, an empty localized Ce\,$4f$ manifold at the conduction edge, and Ce\,$5d$ states ${\sim}2$\,eV above; this level ordering is obtained at every level of theory applied here, with the gaps of Table~\ref{tab:bulk}. Pure PBE places the empty $4f$ band only ${\sim}1.9$\,eV above the valence edge, the self-interaction error characteristic of localized $4f$ states in GGA~\cite{Skorodumova2002, Loschen2007}; the linear-response $U$ (Sec.~\ref{subsec:dft}) opens the gap to $2.35$\,eV, and HSE06~\cite{Krukau2006} recovers $3.21$\,eV, close to the experimental $E_g\approx3.3$\,eV~\cite{Han2016, Wuilloud1984}. The accompanying ${\sim}1.8\%$ overestimate of the lattice parameter is the documented GGA$+U$ behavior for ceria~\cite{Loschen2007} and does not affect the localization physics that follows.

\begin{table}[ht]
	\caption{\label{tab:bulk}Computed and experimental bulk properties of pristine \ce{CeO2}. The HSE06 gap was evaluated at the PBE$+U$ geometry.}
	\begin{ruledtabular}
		\begin{tabular}{lcccc}
			Property    & PBE   & PBE$+U$ & HSE06 & Expt.~\cite{Han2016} \\
			\hline
			$E_g$ (eV)  & 1.86  & 2.35    & 3.21  & 3.3                  \\
			$a_0$ (\AA) & 5.464 & 5.507   & ---   & 5.41                 \\
		\end{tabular}
	\end{ruledtabular}
\end{table}

The narrow (${\sim}1.5$\,eV) empty $4f$ band is the signature of the strong $4f$ localization that drives polaron formation on electron donation: any in-gap state appearing upon vacancy creation occupies one or two specific Ce\,$4f$ orbitals rather than a delocalized band state. The $2\times2\times1$ supercell reproduces the primitive-cell gap and valence-band width to better than $0.05$\,eV, so defect states are referenced cleanly against the pristine supercell. We use PBE$+U$ for all defect calculations and HSE06 only as a gap benchmark.

\subsection{\CeIII{} Polaron Formation and Magnetic Order}\label{subsec:charge}
\begin{figure*}
	\centering
	\begin{subfigure}{.245\textwidth}
		\includegraphics[width=\textwidth]{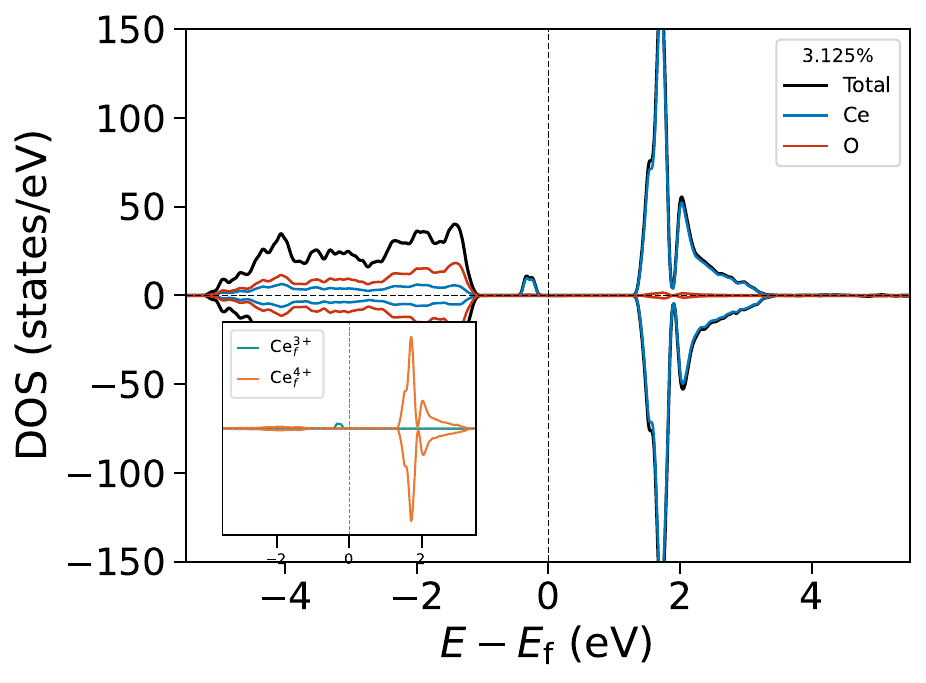}
		\caption{Single, 3.125\%}\label{fig:dos-single}
	\end{subfigure}
	\begin{subfigure}{.24\textwidth}
		\includegraphics[width=\textwidth]{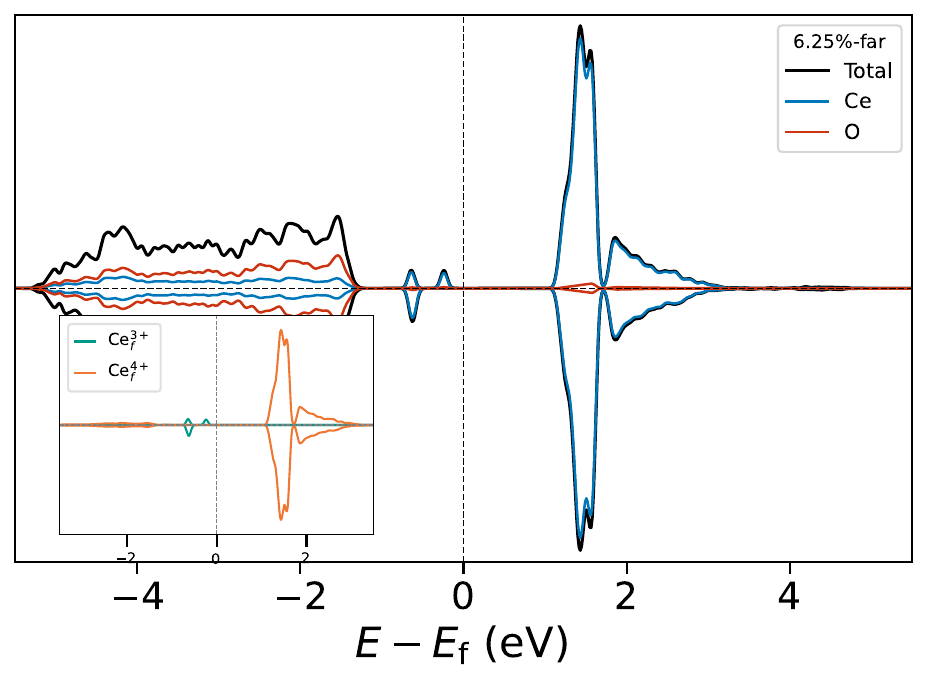}
		\caption{Double-far, 6.25\%}\label{fig:dos-dblfar}
	\end{subfigure}
	\begin{subfigure}{.24\textwidth}
		\includegraphics[width=\textwidth]{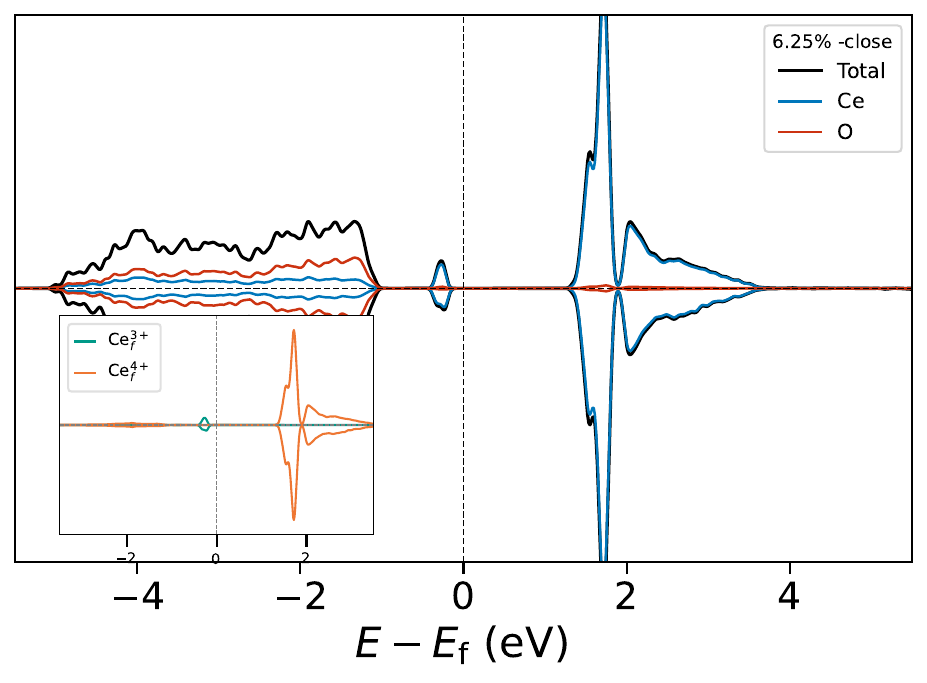}
		\caption{Double-close, 6.25\%}\label{fig:dos-dblclose}
	\end{subfigure}
	\begin{subfigure}{.24\textwidth}
		\includegraphics[width=\textwidth]{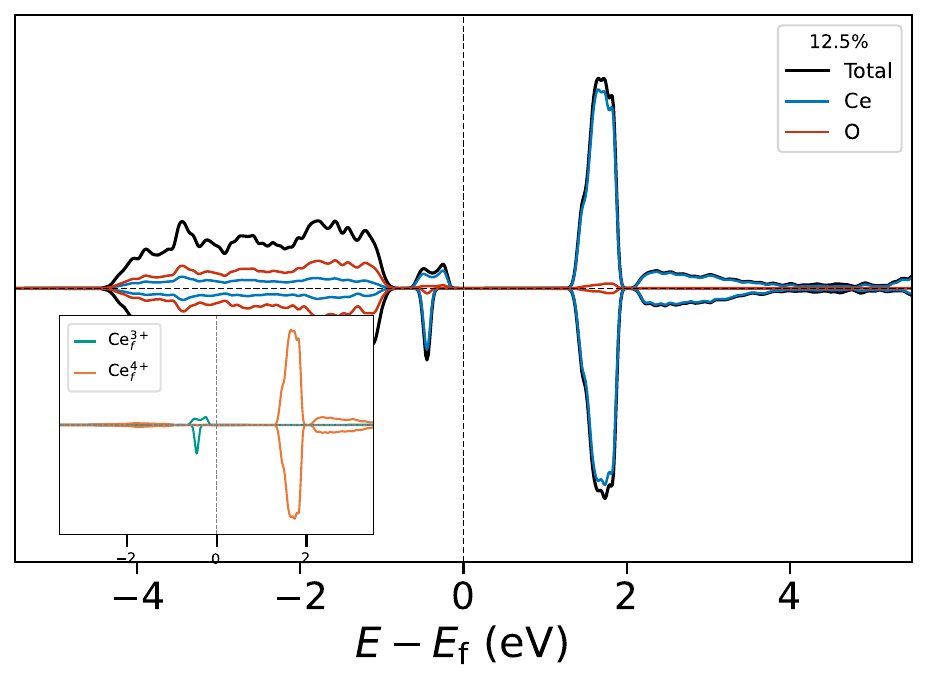}
		\caption{Quad-closer, 12.5\%}\label{fig:dos-quad}
	\end{subfigure}
	\caption{\label{fig:dos}Spin-resolved density of states for the four vacancy configurations, referenced to the Fermi level. Main panels: total, Ce, and O projections (spin-up above, spin-down below the axis); each cell shows an occupied Ce\,$4f$ in-gap state below $E_f$, separated from the O\,$2p$ valence band and the empty Ce\,$4f$ conduction manifold. Insets: $4f$ density of states summed over the \ce{Ce^3+} polaron sites versus the \ce{Ce^4+} lattice sites; the occupied in-gap state is purely \ce{Ce^3+}.}
\end{figure*}

Each oxygen vacancy donates two electrons that localize as two \ce{Ce^3+} ($4f^1$) small polarons, consistent with the experimental identification of \ce{Ce^3+} polarons in reduced ceria~\cite{Jerratsch2011}. We identify polarons by their projected $4f$ moment ($|m_f|=0.96$--$0.97\,\mu_B$, against a \ce{Ce^4+} background $|m_f|\lesssim0.01\,\mu_B$; configuration-averaged values in Table~\ref{tab:moments}, Appendix~\ref{app:mom}); in every cell the number of \ce{Ce^3+} centers is twice the number of vacancies (Table~\ref{tab:magnetic}).In the relaxed geometries the polarons occupy nearest-neighbor cerium sites relative to their vacancy, with average \ce{Ce^3+}--vacancy distances of $2.42$--$2.56$\,\AA{} across the four configurations---a modest ($1$--$8\%$) expansion from the ideal fluorite Ce--O bond length of $2.384$\,\AA{}, consistent with the larger ionic radius of \ce{Ce^3+} relative to \ce{Ce^4+}~\cite{Shannon1976}, and in agreement with the near-vacancy polaron siting reported for bulk ceria~\cite{Murgida2014, Ganduglia2009}. The localization is directly visible in the density of states (Fig.~\ref{fig:dos}): each vacancy adds an occupied Ce\,$4f$ peak $0.3$--$0.6$\,eV below the Fermi level, cleanly split from the O\,$2p$ valence band and the empty Ce\,$4f$ conduction manifold. The four configurations span the dilute single-vacancy limit ($3.125\%$, one isolated polaron pair) to progressively interacting arrangements (up to $12.5\%$), so both the isolated-center regime and its breakdown are accessible within one series.

\begin{table}[h]
	\caption{\label{tab:magnetic}Spin configuration of each vacancy system. $N_{\mathrm{Ce}^{3+}}$ is the number of polaronic \ce{Ce^3+} centers, $n_\uparrow/n_\downarrow$ the majority/minority counts, and $S_\mathrm{net}=\tfrac12(n_\uparrow-n_\downarrow)$. ``Order'' denotes the spin arrangement of the converged self-consistent solution (FM, ferromagnetic; comp., spin-compensated); the multi-vacancy compensated solutions are near-degenerate (Sec.~\ref{subsec:exchange}).}
	\begin{ruledtabular}
		\begin{tabular}{lccccc}
			Configuration & Conc.\ (\%) & $N_{\mathrm{Ce}^{3+}}$ & $n_\uparrow/n_\downarrow$ & $S_\mathrm{net}$ & Order \\
			\hline
			Single        & 3.125       & 2                      & 2/0                       & 1                & FM    \\
			Double-far    & 6.25        & 4                      & 2/2                       & 0                & comp. \\
			Double-close  & 6.25        & 4                      & 2/2                       & 0                & comp. \\
			Quad-closer   & 12.5        & 8                      & 4/4                       & 0                & comp. \\
		\end{tabular}
	\end{ruledtabular}
\end{table}

The single vacancy ($3.125\%$) places two polarons on Ce(4) and Ce(15), coupled ferromagnetically ($S_\mathrm{net}=1$). Every multi-vacancy cell instead converges to a fully compensated solution ($S_\mathrm{net}=0$): the polarons occupy well-separated Ce sites with alternating moment sign, and the moment magnitude is saturated in all cases including the clustered $12.5\%$ cell. We treat each polaron as an effective $S=1/2$ moment, the low-energy label for the \ce{Ce^3+} $4f^1$ Kramers state; the spin--orbit and crystal-field splitting of that manifold, which set the anisotropy of the effective $g$-tensor, are not resolved by the present collinear calculations and are the subject of a companion all-electron study, but they do not affect the transferred spin density that determines the \ce{^17O} hyperfine couplings reported below.

\subsection{Inter-Polaron Exchange Coupling}\label{subsec:exchange}
\begin{figure*}
	\centering
	\begin{subfigure}{.32\textwidth}
		\includegraphics[width=\textwidth]{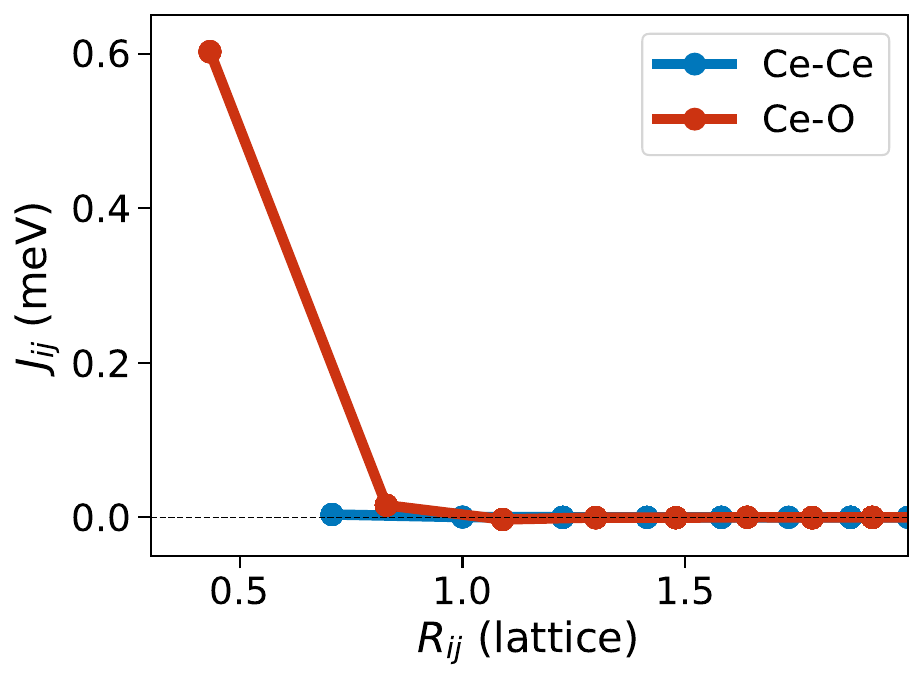}
		\caption{3.125\%}\label{fig:jij-1}
	\end{subfigure}\hfill
	\begin{subfigure}{.32\textwidth}
		\includegraphics[width=\textwidth]{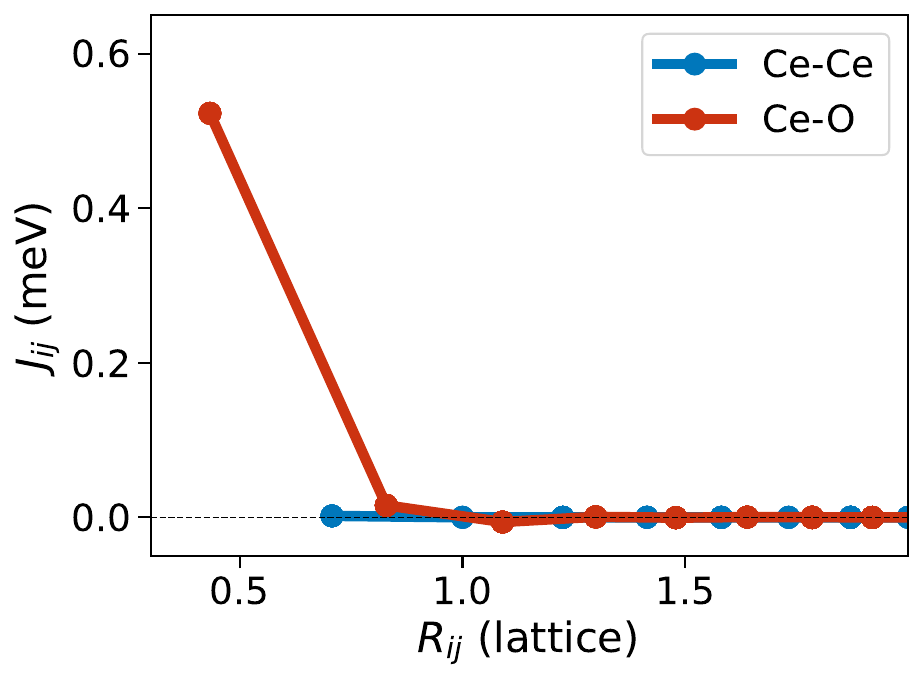}
		\caption{6.25\%}\label{fig:jij-2}
	\end{subfigure}\hfill
	\begin{subfigure}{.32\textwidth}
		\includegraphics[width=\textwidth]{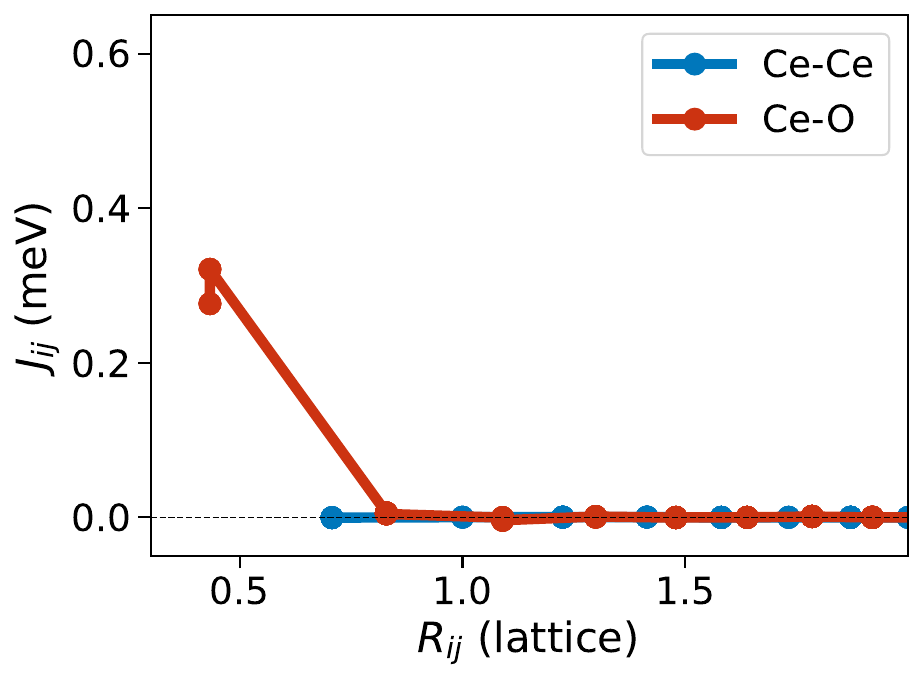}
		\caption{12.5\%}\label{fig:jij-4}
	\end{subfigure}
	\caption{\label{fig:jij}KKR-CPA exchange constants $J_{ij}$ versus pair separation $R_{ij}$ (in lattice units) for vacancy concentrations of (a)~3.125\%, (b)~6.25\%, and (c)~12.5\%. \ce{Ce}--\ce{Ce} (blue) and \ce{Ce}--\ce{O} (orange) couplings are shown; $J_{ij}>0$ is ferromagnetic. A single ferromagnetic nearest-neighbor \ce{Ce}--\ce{O} term dominates and decreases with concentration, while direct \ce{Ce}--\ce{Ce} exchange is negligible at all separations.}
\end{figure*}

To characterize the coupling between polarons independently of the particular supercell spin solution, we computed Heisenberg exchange parameters $J_{ij}$ within all-electron KKR-CPA (Sec.~\ref{subsec:sprkkr}), treating the vacancy as a fractional component on the anion sublattice; as discussed there, these $J_{ij}$ are configuration- and valence-averaged, dilute-limit estimates of the exchange scale, complementary to the site-resolved supercell solutions. $J_{ij}>0$ is ferromagnetic. The direct \ce{Ce}--\ce{Ce} exchange is negligible at every concentration ($|J|\leq4\times10^{-3}$\,meV), confirming that the polarons do not couple through direct $4f$--$4f$ overlap. The coupling is carried entirely by the oxygen sublattice: the dominant term is the nearest-neighbor \ce{Ce}--\ce{O} exchange at $R=0.43\,a_0$, which is ferromagnetic and decreases monotonically with concentration, from $+0.60$\,meV at $3.125\%$ to $+0.30$\,meV at $12.5\%$ (Fig.~\ref{fig:jij}, Table~\ref{tab:jij}); higher shells are one to two orders of magnitude weaker and alternate in sign. The same oxygen sublattice carries the entire \ce{^17O} hyperfine response of Sec.~\ref{subsec:hfc}, so the magnetic and hyperfine channels share the \ce{Ce}--\ce{O}--\ce{Ce} superexchange pathway.

\begin{table}[h]
	\caption{\label{tab:jij}Dominant KKR-CPA exchange constants by neighbor shell and type. $R$ is in units of $a_0$; $J>0$ is ferromagnetic. The $6.25\%$ column is the CPA composition average and does not distinguish the close and far geometries.}
	\begin{ruledtabular}
		\begin{tabular}{lcddd}
			Pair                          & $R/a_0$ &
			\multicolumn{1}{c}{$3.125\%$} &
			\multicolumn{1}{c}{$6.25\%$}  &
			\multicolumn{1}{c}{$12.5\%$}                                                                   \\
			                              &         & \multicolumn{3}{c}{$J_{ij}$ (meV)}                   \\
			\hline
			\ce{Ce}--\ce{O}               & 0.43    & +0.602                             & +0.523 & +0.299 \\
			\ce{Ce}--\ce{O}               & 0.83    & +0.016                             & +0.015 & +0.005 \\
			\ce{Ce}--\ce{O}               & 1.09    & -0.003                             & -0.006 & -0.002 \\
			\ce{Ce}--\ce{Ce}              & 0.71    & +0.004                             & +0.002 & -0.000 \\
			\ce{Ce}--\ce{Vc}              & 0.43    & -0.002                             & +0.000 & +0.016 \\
		\end{tabular}
	\end{ruledtabular}
\end{table}

The two probes are complementary by construction and converge on a single physical picture. The CPA returns the disorder-averaged pairwise exchange of a dilute vacancy ensemble: ferromagnetic in its dominant first-neighbor \ce{Ce}--\ce{O} term and sub-meV in magnitude at every concentration. Each supercell instead realizes one specific ordered geometry, and its converged spin state samples a configuration landscape whose candidate states all lie within that same sub-meV window: the single-vacancy cell adopts the ferromagnetic alignment favored by the dominant coupling, while the multi-vacancy cells settle into compensated arrangements near-degenerate with it (Table~\ref{tab:magnetic}). This near-degeneracy is itself the central magnetic result: with the full exchange scale at $0.3$--$0.6$\,meV---a few kelvin---the spin-configuration landscape is essentially flat at any temperature relevant to spin-center operation, and the particular arrangement a given cell adopts has no bearing on the per-polaron quantities that transfer to the isolated-center limit. On every quantity that does matter, the two methods agree: the coupling is weak, short-ranged and dominated by the first \ce{Ce}--\ce{O} shell, and carried entirely by the oxygen sublattice, with no resolvable direct $4f$--$4f$ channel, so the \ce{Ce^3+} polarons are magnetically dilute and behave as near-independent $S=1/2$ moments.
% The dominant coupling is ferromagnetic, consistent with the ferromagnetic single-vacancy solution but in apparent tension with the compensated multi-vacancy solutions (Table~\ref{tab:magnetic}). This is not frustration---a net-ferromagnetic, short-ranged coupling is not frustrated---but a difference of limits: KKR-CPA returns the configurationally averaged pairwise exchange of a dilute vacancy distribution, whereas the supercells are specific clustered geometries whose candidate spin states, all lying within the sub-meV exchange window, are near-degenerate; assigning the clustered ground state unambiguously would require an explicit total-energy comparison of those states, which we do not attempt here. The conclusion that matters for the spin centers is independent of that assignment: the polarons couple only weakly, only through the oxygen sublattice, and only at first-neighbor range. With a dominant exchange of order $0.3$--$0.6$\,meV---an energy scale of a few kelvin---the \ce{Ce^3+} centers are magnetically dilute and behave as near-independent $S=1/2$ moments at any temperature relevant to spin-center operation.

\subsection{Hyperfine Structure: \ce{^17O} as the Sole Coupling Channel}\label{subsec:hfc}
\subsubsection{Absence of cerium hyperfine coupling}

A defining feature of \CeIII{}:\CeO{} follows directly from cerium isotopics: every stable Ce isotope has nuclear spin $I=0$~\cite{Stone2005}, so the \ce{Ce^3+} electron spin has no on-site hyperfine partner. This is an exact consequence of the vanishing nuclear moment, not a computed result; what the calculation quantifies is its consequence. The $4f^1$ spin density is fully localized on the polaron site (Sec.~\ref{subsec:charge}), so the interaction that dominates the spin Hamiltonian of essentially every conventional paramagnetic center---the coupling to the largest on-site spin density---is silent here, and the entire hyperfine budget of the defect is carried by the small spin density transferred to the ligands. The host cation therefore contributes no nuclear-spin decoherence channel, and the hyperfine environment resides exclusively on oxygen, whose only magnetic isotope is \ce{^17O} ($I=5/2$) at $0.038\%$ natural abundance~\cite{Meija2016}. The resulting bath is exceptionally sparse---one to two orders of magnitude below the \ce{^13C} bath of diamond ($1.1\%$) or the \ce{^29Si} bath of silicon carbide ($4.7\%$)---and tunable by \ce{^17O} depletion or enrichment.
It differs from those spin-$1/2$ baths in one further respect: \ce{^17O} carries an electric quadrupole moment~\cite{Stone2005}, so its coupling to the local electric-field gradient splits the nuclear Zeeman ladder site by site, detuning energy-conserving flip-flops between crystallographically inequivalent nuclei and thereby expected to further suppress the spectral diffusion that a bath of equal density would otherwise produce.

\subsubsection{\ce{^17O} contact couplings}

Table~\ref{tab:O17} lists the most strongly coupled \ce{^17O} sites in each configuration; the full per-site tensors are given in Appendix~\ref{app:hyp}. A common structure emerges: each polaron is coordinated by a few oxygen sites carrying a large,
nearly axial contact coupling ($|A_\mathrm{iso}|$ up to $5.9$\,MHz, $|\eta|<0.1$)---the first-shell ligands that overlap the $4f$
spin density directly---surrounded by a shell of order-of-magnitude weaker, strongly anisotropic couplings. These strong
first-shell couplings are short-range and set by spin density within the first coordination sphere, the regime least sensitive to
periodic-image error (Sec.~\ref{subsec:hfmethod}). Beyond these shells the transferred spin density is negligible and the
coupling reduces to the analytic point dipole of the polaron moment---the regime where the periodic-image contamination of the
DFT values is irrelevant because the DFT values are not needed---so the ab initio first-shell set together with the dipolar tail
constitutes the complete bath parametrization that a cluster-correlation-expansion $T_2$ calculation requires~\cite{Yang2008,
	Onizhuk2025}.
\begin{table}[h]
	\caption{\label{tab:O17}Most strongly coupled \ce{^17O} sites in each configuration, including the core-polarization contribution $A_{1c}$: $A_\mathrm{iso}$, largest principal value $A_{zz}$, and asymmetry $\eta$ ($|A_{zz}|\geq|A_{xx}|\geq|A_{yy}|$). Representative strong (axial) and weak (anisotropic) sites are shown; full tables in Appendix~\ref{app:hyp}. Signs track the polaron sublattice.}
	\begin{ruledtabular}
		\begin{tabular}{lddd}
			Site  & \multicolumn{1}{c}{$A_\mathrm{iso}$} & \multicolumn{1}{c}{$A_{zz}$} & \multicolumn{1}{c}{$\eta$} \\
			      & \multicolumn{2}{c}{(MHz)}            &                                                           \\
			\hline
			\multicolumn{4}{l}{\textit{3.125\% single (FM)}}                                                         \\
			O(37) & 2.171                                & 2.518                        & -0.075                     \\
			O(38) & 2.157                                & 2.503                        & -0.076                     \\
			O(30) & 1.080                                & 1.705                        & -0.026                     \\
			O(22) & 0.666                                & 1.082                        & -0.843                     \\
			O(46) & 0.666                                & 1.081                        & -0.841                     \\
			\hline
			\multicolumn{4}{l}{\textit{6.25\% double-far (compensated)}}                                             \\
			O(38) & 5.907                                & 7.600                        & -0.082                     \\
			O(35) & -4.730                               & -5.505                       & -0.103                     \\
			O(36) & -4.528                               & -5.255                       & -0.109                     \\
			O(37) & 4.453                                & 5.131                        & -0.113                     \\
			O(45) & 1.439                                & 2.323                        & -0.206                     \\
			\hline
			\multicolumn{4}{l}{\textit{6.25\% double-close (compensated)}}                                           \\
			O(35) & 4.745                                & 6.012                        & -0.111                     \\
			O(41) & 4.735                                & 6.013                        & -0.109                     \\
			O(36) & -4.720                               & -5.970                       & -0.113                     \\
			O(38) & -4.622                               & -5.859                       & -0.114                     \\
			O(27) & -3.317                               & -4.998                       & -0.473                     \\
			\hline
			\multicolumn{4}{l}{\textit{12.5\% quad-closer (compensated)}}                                            \\
			O(28) & 5.264                                & 5.936                        & -0.032                     \\
			O(29) & 5.250                                & 5.909                        & -0.030                     \\
			O(37) & 5.048                                & 5.675                        & -0.048                     \\
			O(36) & 4.625                                & 5.366                        & -0.053                     \\
			O(30) & -2.628                               & -3.522                       & -0.035                     \\
		\end{tabular}
	\end{ruledtabular}
\end{table}
The signs of the couplings track the polaron sublattice: in the ferromagnetic single-vacancy cell all appreciable
couplings share one sign, while in the compensated cells they appear in opposite-sign groups, with oxygens bridging antiparallel
polarons sampling partially canceling spin density. As concentration rises, the polaron coordination shells begin to overlap---at
$12.5\%$ every oxygen in the cell carries a coupling above $0.5$\,MHz---which is the microscopic form of the isolated-center
breakdown anticipated above.

\subsubsection{Anisotropy and the \ce{^17O} target pattern}\label{subsec:17O}

The asymmetry $\eta$ is an intra-site ratio of principal values and is comparatively robust to finite-size effects that scale all
three values together. The near-polaron sites are uniformly near-axial ($|\eta|\lesssim0.1$), the signature of a
contact-dominated coupling along the \ce{Ce}--\ce{O} bond, while the weaker shell sites are dipolar-dominated and strongly
anisotropic. Among the sites with well-defined tensors ($|A_\mathrm{iso}|\geq0.5$\,MHz; below this the coupling is negligible and
$\eta$ is numerically ill-conditioned), the shell asymmetries of the two $6.25\%$ geometries span overlapping ranges
($|\eta|\approx0.47$--$0.70$ close, $0.15$--$0.72$ far), so $\eta$ alone does not cleanly discriminate clustered from dispersed
arrangements. We therefore present the \ce{^17O} tensors not as a geometric ruler but as a first-principles target pattern for
pulsed EPR. The predicted couplings fall in the window these techniques exploit: at X band the \ce{^17O} Larmor frequency is
${\sim}2$\,MHz, so the first-shell couplings lie in the $|A|\sim2\nu_I$ regime of deep envelope modulation~\cite{Mims1972,
	Schweiger2001}, and the two-tier hierarchy should appear in HYSCORE as strong-coupling cross-peaks from the first-shell sites,
resolved from the weak-coupling ridge of the anisotropic shell near the \ce{^17O} Larmor position~\cite{Hoefer1986,
	Schweiger2001}. One ingredient beyond the present tensors enters a quantitative $I=5/2$ simulation: the nuclear-quadrupole
couplings set by the electric-field gradients at the oxygen sites, which shift the modulation frequencies alongside the hyperfine
terms. These do not alter the tensors reported here and are directly computable within the same PAW framework; the tabulated
$\{A_\mathrm{iso}, A_{zz}, \eta\}$ fix the hyperfine part of the pattern that a \ce{^17O}-enriched reduced-ceria sample should
display.

\subsection{Vacancy Formation Energies}\label{subsec:Ef}
Oxygen-vacancy formation energies in the oxygen-rich limit are reported in Table~\ref{tab:Ef}. As discussed in Sec.~\ref{subsec:dft}, the GGA$+U$ molecular reference shifts all values by a common offset, so the absolute magnitudes are not directly comparable to values computed with corrected oxygen references, and only differences within the series are interpreted. Two comparisons follow. At fixed $6.25\%$ the dispersed (double-far) arrangement is favored over the clustered (double-close) by $0.34$\,eV per vacancy, and this preference is independent of the compensated spin state both geometries adopt. Along the clustered branch the per-vacancy cost decreases with concentration, from $7.79$\,eV at $3.125\%$ to $6.79$\,eV at $12.5\%$, indicating that vacancy formation is cooperative rather than penalized, consistent with vacancy--vacancy attraction in reduced ceria~\cite{Murgida2014, Ganduglia2009}. For the spin centers the practical implication is that neither the dilute isolated-center regime nor the clustered multi-center regime is thermodynamically inaccessible.

\begin{table}[h]
	\caption{\label{tab:Ef}Oxygen-vacancy formation energies per vacancy, oxygen-rich limit. The molecular-reference offset is common to all rows (Sec.~\ref{subsec:dft}); only differences are interpreted.}
	\begin{ruledtabular}
		\begin{tabular}{lcc}
			Configuration & Conc.\ (\%) & $E_\mathrm{f}$ (eV/vac.) \\
			\hline
			Single        & 3.125       & 7.792                    \\
			Double-far    & 6.25        & 7.476                    \\
			Double-close  & 6.25        & 7.817                    \\
			Quad-closer   & 12.5        & 6.790                    \\
		\end{tabular}
	\end{ruledtabular}
\end{table}

\section{Conclusions and Outlook}
\label{sec:conclusion}

We have mapped the hyperfine interactions, magnetic order, and inter-polaron exchange of oxygen-vacancy-induced \ce{Ce^3+} spin centers in \ce{CeO2} from first principles, across four configurations spanning the dilute single-vacancy limit to the interacting $12.5\%$ regime. The defining feature of the host follows from cerium isotopics: every stable cerium isotope has $I=0$, so the element that carries the electron spin is itself nuclear-spin-free---the center and the spin-free sublattice are the same chemical species---and the entire hyperfine environment falls to \ce{^17O} at $0.038\%$ natural abundance, a bath one to two orders of magnitude sparser than the \ce{^13C} and \ce{^29Si} baths of diamond and silicon carbide, obtained from native stoichiometry rather than isotopic purification.

The \ce{^17O} landscape is two-tiered---a few strong, nearly axial first-shell couplings reaching ${\sim}5.9$\,MHz over a weakly coupled, strongly anisotropic outer shell---and, joined to the analytic point-dipole tail that holds beyond the range of transferred spin density, it constitutes a complete bath parametrization for cluster-correlation-expansion calculations of $T_2$. The prediction is directly testable: at X band the first-shell couplings lie in the deep-modulation window of \ce{^17O} ESEEM, and HYSCORE of a \ce{^17O}-enriched reduced-ceria sample should resolve the two tiers as distinct strong- and weak-coupling features. The exchange between polarons is weak (sub-meV), carried entirely by the oxygen sublattice with no resolvable direct $4f$--$4f$ channel, so the centers behave as near-independent $S=1/2$ moments at any temperature relevant to operation; the quadrupolar $I=5/2$ character of the bath, whose site-dependent splittings detune nuclear flip-flops, should suppress spectral diffusion below what its density alone would imply.

The host thus offers two independent control axes: the bath strength, set by the \ce{^17O} fraction and tunable by depletion or enrichment, and the inter-center spacing and coupling, set by vacancy geometry, with dispersed and clustered arrangements both thermodynamically accessible. This hyperfine-and-exchange characterization is the first layer of a full $\{A, J, g, T_2\}$ description of \ce{Ce^3+}:\ce{CeO2}; the $g$-tensor of the $^2F_{5/2}$-derived Kramers doublet and the coherence times that follow from the bath mapped here are the subjects of companion studies, and optical initialization and readout of the center remain undemonstrated. What is established is the design space itself: a rare-earth Kramers spin generated by controlled reduction rather than implantation, whose spacing and coupling are set by vacancy geometry, in a host whose spin-bearing element is intrinsically free of nuclear spin---a conjunction unavailable to implanted point defects and extrinsic dopants, and one whose static magnetic parameters are now fixed.
\appendix

\section{Ce\,$4f$ Magnetic Moments}\label{app:mom}

Table~\ref{tab:moments} reports the projected Ce\,$4f$ moment averaged separately over the polaronic \ce{Ce^3+} sites ($|m_f|>0.5\,\mu_B$) and the \ce{Ce^4+} lattice sites, for each configuration. The near-integer polaron moments and the near-vanishing lattice-site moments (below $0.01\,\mu_B$ in every cell) confirm clean, complete charge localization with no measurable delocalization onto the nominally \ce{Ce^4+} sublattice.

\begin{table}[h]
	\caption{\label{tab:moments}Projected Ce\,$4f$ magnetic moments, averaged over the polaronic \ce{Ce^3+} sites and separately over the \ce{Ce^4+} lattice sites, for each configuration. $N_{\mathrm{Ce}^{3+}}$ is the number of polaron sites (cf.\ Table~\ref{tab:magnetic}).}
	\begin{ruledtabular}
		\begin{tabular}{lcccc}
			Configuration & $N_{\mathrm{Ce}^{3+}}$ & $\langle|m_f|\rangle_{\mathrm{Ce}^{3+}}$ & $\langle|m_f|\rangle_{\mathrm{Ce}^{4+}}$ & max.\ \ce{Ce^4+} \\
			              &                        & ($\mu_B$)                                & ($\mu_B$)                                & ($\mu_B$)        \\
			\hline
			Single        & 2                      & 0.974                                    & 0.001                                    & 0.003            \\
			Double-far    & 4                      & 0.968                                    & 0.003                                    & 0.007            \\
			Double-close  & 4                      & 0.964                                    & 0.003                                    & 0.004            \\
			Quad-closer   & 8                      & 0.971                                    & 0.004                                    & 0.007            \\
		\end{tabular}
	\end{ruledtabular}
\end{table}

\section{\ce{^17O} Hyperfine Tensors}\label{app:hyp}

The \ce{Ce^3+} centers carry no hyperfine coupling (Sec.~\ref{subsec:hfc}); the complete hyperfine response of each cell resides
on the oxygen sublattice. Tables~\ref{tab:O17full_s}--\ref{tab:O17full_q} list, for each configuration, the \ce{^17O} sites with
isotropic contact coupling $|A_\mathrm{iso}|\geq0.5$\,MHz, giving $A_\mathrm{iso}$, the largest diagonalized principal value
$A_{zz}$ (convention $|A_{zz}|\geq|A_{xx}|\geq|A_{yy}|$), and the asymmetry $\eta=(A_{yy}-A_{xx})/A_{zz}$, listed by \ce{O}-site
index. These are the short-range couplings set by spin density in the first oxygen shells and are the most strongly converged hyperfine quantities. The remaining oxygen sites in each cell carry negligible contact coupling ($|A_\mathrm{iso}|<0.5$\,MHz: 18 of 31 sites for single, 18 of 30 for double-close, 9 of 30 for double-far, and 0 of 28 for quad-closer); their absolute couplings are additionally subject to the periodic-image error of Ref.~\cite{Takacs2024} (Sec.~\ref{subsec:hfmethod}) and their $\eta$ is numerically ill-conditioned, so they are omitted. Signs of $A_\mathrm{iso}$ track the local polaron sublattice; in the compensated multi-vacancy cells the couplings therefore appear in opposite-sign groups. Across the two $6.25\%$ geometries the near-polaron sites are uniformly near-axial and the shell sites span overlapping $|\eta|$ ranges (Sec.~\ref{subsec:17O}).

\begin{table}[h]
	\caption{\label{tab:O17full_s}\ce{^17O} hyperfine tensors, 3.125\% single vacancy (FM). Contact term includes the core-polarization contribution $A_{1c}$; sites listed by O index.}
	\begin{ruledtabular}
		\begin{tabular}{lddd}
			Site  & \multicolumn{1}{c}{$A_\mathrm{iso}$} & \multicolumn{1}{c}{$A_{zz}$} & \multicolumn{1}{c}{$\eta$} \\
			      & \multicolumn{2}{c}{(MHz)}            &                                                           \\
			\hline
			O(22) & 0.666                                & 1.082                        & -0.843                     \\
			O(23) & 0.662                                & 1.076                        & -0.845                     \\
			O(26) & 0.662                                & 0.929                        & -0.752                     \\
			O(27) & 0.655                                & 0.921                        & -0.754                     \\
			O(30) & 1.080                                & 1.705                        & -0.026                     \\
			O(32) & 0.654                                & 0.920                        & -0.754                     \\
			O(34) & 0.663                                & 0.930                        & -0.749                     \\
			O(37) & 2.171                                & 2.518                        & -0.075                     \\
			O(38) & 2.157                                & 2.503                        & -0.076                     \\
			O(41) & 0.618                                & 1.067                        & -0.748                     \\
			O(42) & 0.617                                & 1.067                        & -0.747                     \\
			O(44) & 0.663                                & 1.078                        & -0.844                     \\
			O(46) & 0.666                                & 1.081                        & -0.841                     \\
		\end{tabular}
	\end{ruledtabular}
\end{table}

\begin{table}[h]
	\caption{\label{tab:O17full_df}\ce{^17O} hyperfine tensors, 6.25\% double-far (compensated), including $A_{1c}$; sites listed by O index. Opposite signs label the two polaron sublattices.}
	\begin{ruledtabular}
		\begin{tabular}{lddd}
			Site  & \multicolumn{1}{c}{$A_\mathrm{iso}$} & \multicolumn{1}{c}{$A_{zz}$} & \multicolumn{1}{c}{$\eta$} \\
			      & \multicolumn{2}{c}{(MHz)}            &                                                           \\
			\hline
			O(19) & 1.325                                & 2.305                        & -0.639                     \\
			O(20) & -1.350                               & -2.327                       & -0.639                     \\
			O(21) & -1.323                               & -2.285                       & -0.645                     \\
			O(22) & 1.359                                & 1.826                        & -0.153                     \\
			O(23) & 1.245                                & 1.839                        & -0.588                     \\
			O(24) & -1.229                               & -1.827                       & -0.608                     \\
			O(25) & -1.174                               & -1.726                       & -0.638                     \\
			O(31) & -1.178                               & -2.180                       & -0.719                     \\
			O(32) & -1.249                               & -2.310                       & -0.677                     \\
			O(34) & 1.265                                & 2.322                        & -0.682                     \\
			O(35) & -4.730                               & -5.505                       & -0.103                     \\
			O(36) & -4.528                               & -5.255                       & -0.109                     \\
			O(37) & 4.453                                & 5.131                        & -0.113                     \\
			O(38) & 5.907                                & 7.600                        & -0.082                     \\
			O(39) & -1.278                               & -2.251                       & -0.671                     \\
			O(40) & -1.223                               & -2.204                       & -0.642                     \\
			O(41) & 1.350                                & 2.327                        & -0.681                     \\
			O(43) & -1.423                               & -2.804                       & -0.643                     \\
			O(44) & -1.402                               & -2.745                       & -0.642                     \\
			O(45) & 1.439                                & 2.323                        & -0.206                     \\
			O(46) & 1.356                                & 2.695                        & -0.646                     \\
		\end{tabular}
	\end{ruledtabular}
\end{table}

\begin{table}[h]
	\caption{\label{tab:O17full_dc}\ce{^17O} hyperfine tensors, 6.25\% double-close (compensated), including $A_{1c}$; sites listed by O index. Opposite signs label the two polaron sublattices.}
	\begin{ruledtabular}
		\begin{tabular}{lddd}
			Site  & \multicolumn{1}{c}{$A_\mathrm{iso}$} & \multicolumn{1}{c}{$A_{zz}$} & \multicolumn{1}{c}{$\eta$} \\
			      & \multicolumn{2}{c}{(MHz)}            &                                                           \\
			\hline
			O(22) & 3.131                                & 4.664                        & -0.524                     \\
			O(23) & -2.931                               & -4.191                       & -0.603                     \\
			O(26) & 3.076                                & 4.552                        & -0.536                     \\
			O(27) & -3.317                               & -4.998                       & -0.473                     \\
			O(35) & 4.745                                & 6.012                        & -0.111                     \\
			O(36) & -4.720                               & -5.970                       & -0.113                     \\
			O(37) & 1.706                                & 2.915                        & -0.698                     \\
			O(38) & -4.622                               & -5.859                       & -0.114                     \\
			O(39) & 1.718                                & 2.910                        & -0.688                     \\
			O(40) & -1.766                               & -2.953                       & -0.667                     \\
			O(41) & 4.735                                & 6.013                        & -0.109                     \\
			O(42) & -1.868                               & -3.028                       & -0.636                     \\
		\end{tabular}
	\end{ruledtabular}
\end{table}

\begin{table}[h]
	\caption{\label{tab:O17full_q}\ce{^17O} hyperfine tensors, 12.5\% quad-closer (compensated), including $A_{1c}$; sites listed by O index. Opposite signs label the two polaron sublattices. All 28 oxygen sites exceed the $0.5$\,MHz cutoff.}
	\begin{ruledtabular}
		\begin{tabular}{lddd}
			Site  & \multicolumn{1}{c}{$A_\mathrm{iso}$} & \multicolumn{1}{c}{$A_{zz}$} & \multicolumn{1}{c}{$\eta$} \\
			      & \multicolumn{2}{c}{(MHz)}            &                                                           \\
			\hline
			O(17) & -1.273                               & -1.839                       & -0.647                     \\
			O(18) & -1.290                               & -1.851                       & -0.650                     \\
			O(19) & 0.685                                & 1.732                        & -0.175                     \\
			O(20) & 0.571                                & 1.789                        & +0.216                     \\
			O(21) & 0.577                                & 1.615                        & -0.174                     \\
			O(22) & 0.704                                & 1.601                        & -0.142                     \\
			O(23) & -1.563                               & -2.835                       & -0.397                     \\
			O(24) & -1.286                               & -2.563                       & -0.449                     \\
			O(25) & -1.554                               & -2.819                       & -0.403                     \\
			O(26) & -1.554                               & -2.768                       & -0.398                     \\
			O(27) & -2.589                               & -3.498                       & -0.031                     \\
			O(28) & 5.264                                & 5.936                        & -0.032                     \\
			O(29) & 5.250                                & 5.909                        & -0.030                     \\
			O(30) & -2.628                               & -3.522                       & -0.035                     \\
			O(31) & -1.318                               & -2.592                       & -0.451                     \\
			O(32) & -1.509                               & -2.712                       & -0.399                     \\
			O(33) & -1.574                               & -2.853                       & -0.393                     \\
			O(34) & -1.598                               & -2.879                       & -0.414                     \\
			O(35) & -2.538                               & -3.387                       & -0.051                     \\
			O(36) & 4.625                                & 5.366                        & -0.053                     \\
			O(37) & 5.048                                & 5.675                        & -0.048                     \\
			O(38) & -2.612                               & -3.483                       & -0.046                     \\
			O(39) & -1.240                               & -1.799                       & -0.690                     \\
			O(40) & -1.279                               & -1.836                       & -0.664                     \\
			O(41) & 0.661                                & 1.864                        & -0.193                     \\
			O(42) & 0.719                                & 1.606                        & -0.143                     \\
			O(43) & 0.653                                & 1.706                        & -0.175                     \\
			O(44) & 0.591                                & 1.634                        & -0.171                     \\
		\end{tabular}
	\end{ruledtabular}
\end{table}

\bibliography{refs}

\end{document}